\documentclass[prb,twocolumn,showpacs,preprintnumbers,amsmath,amssymb,superscriptaddress]{revtex4}
\usepackage{epsf}
\usepackage{graphicx}
\usepackage{bm}% bold math

\usepackage{color}

\begin{document}

\title{Conversion of hole states by acoustic solitons}

\author{I.~V.~Rozhansky}
%\email{igor@quantum.ioffe.ru}
\affiliation{A.F.~Ioffe Physico-Technical Institute, Russian Academy of
Sciences, 194021 St.Petersburg, Russia}
\author{M.~B.~Lifshits}
\affiliation{A.F.~Ioffe Physico-Technical Institute, Russian Academy
of Sciences, 194021 St.Petersburg, Russia} \affiliation {Universite
Montpellier II, Montpellier, CEDEX 5, France}
\author{S.~A.~Tarasenko}
\email{tarasenko@coherent.ioffe.ru} 
\affiliation{A.F.~Ioffe Physico-Technical
Institute, Russian Academy of Sciences, 194021 St.Petersburg, Russia}
\author{N.~S.~Averkiev}
\affiliation{A.F.~Ioffe Physico-Technical Institute, Russian Academy
of Sciences, 194021 St.Petersburg, Russia}


%\date{\today}

\begin{abstract}
{The hole states in the valence band of a large class of semiconductors are degenerate in
the projections of angular momentum. Here we show that the switching of a hole
between the states can efficiently be realized by acoustic solitons. The
microscopic mechanism of such a state conversion is related to the valence band splitting 
by local elastic strain. The conversion is studied here for heavy holes localized at shallow 
and deep acceptors in silicon quantum wells.}
\end{abstract}

\maketitle

The strain induced effects have been proved to be an efficient tool to study
basic properties of semiconductor structures yielding information on their
point-group symmetry, band parameters, spin-orbit coupling,
etc.~\cite{BirPikus,Sun07} During the last decade the possibility emerged to
study such effects in nanostructures on the extremely short time scale of
picosecond range. This could be done employing acoustic solitary waves
(solitons).~\cite{Hao01,Muskens04,Scherbakov07,Voronkov99} The acoustic soliton
is a pulse of strong nonlinear elastic strain propagating along the crystal
with the velocity larger than the sound speed.~\cite{Kosevich} Due to rather
high values of deformation-potential constants in semiconductors, such a local
strain efficiently modifies the band structure and thereby interacts with
charge carriers. The important point is that, unlike linear acoustic waves, the
soliton-type strain does not change its sign within the
pulse.~\cite{Hao01,Kosevich} Therefore, after the soliton has passed through
the sample, the electron state of a system may differ from the initial one.

In this paper we show that, in semiconductor structures where the electron or hole
states are degenerate in the absence of strain, acoustic solitons can cause a transition of carriers between these quantum-mechanical states. We study such a state conversion for heavy
holes localized at acceptors in quantum wells (QWs). The ground state
of localized holes in most of the cubic semiconductors is degenerate in the
projection of the angular momentum because of the complex structure of the
valence band. We show that the propagation of the acoustic soliton of a certain
amplitude through the area of hole localization changes the projection of the hole
angular momentum. The effect of state conversion, particularly between two easily distinguished states, is of interest not only from fundamental point of view but can also be utilized for the information processing and storage.

The acoustic soliton represents a perturbation of purely mechanical origin,
therefore, in the first approximation it does not interact with the carrier
spin. Accordingly, it is convenient to consider the effect of state conversion
for semiconductor structures with negligible spin-orbit interaction (such as
Si, SiC, etc). Below we focus on silicon-based quantum wells although the main
conclusions can be generalized to other systems. We assume that in bulk
material the valence-band states at the center of the Brillouin zone belong to
the representations $\Gamma^\prime_{25}$ or $\Gamma_{15}$ and denote the basis
functions as $X$, $Y$, and $Z$. Due to quantum confinement, the ground
heavy-hole states in a structure grown along the $z$ axis are formed from the
Bloch amplitudes $X$ and $Y$.~\cite{Rodriguez99} 
The corresponding wave functions have the form
\begin{equation}\label{Wavefunction}
\Psi(\bm{r}) = \alpha(\bm{\rho}) u(z) X + \beta(\bm{\rho}) u(z) Y \:,
\end{equation}
where $\alpha(\bm{\rho})$ and $\beta(\bm{\rho})$ are smooth envelopes in the
QW plane, which can be combined into a two-component column
$\psi(\bm{\rho})=[\alpha(\bm{\rho}), \beta(\bm{\rho})]^T$,
$\bm{\rho}=(x,y)$ is the in-plane coordinate,
$u(z)$ is the function of size quantization, $x$, $y$, $z$ are the cubic axes, and the spin index is omitted. 

The envelope functions $\psi(\bm{\rho})$ of a hole localized at an acceptor in a narrow quantum well can be found by
solving the matrix Schr\"{o}dinger equation $\hat{H} \psi(\bm{\rho}) = E
\psi(\bm{\rho})$, where $\hat{H}$ is the effective Hamiltonian assuming in the
spherical approximation the form
\begin{equation}
\hat{H} = \gamma \left[ {\begin{array}{*{20}c}
   \bm{k}^2 & {0}  \\
   {0} & \bm{k}^2  \\
\end{array}} \right]+
\gamma^\prime \left[ {\begin{array}{*{20}c}
   {k_x^2  - k_y^2} & 2  {k_x k_y}   \\
   {2 k_x k_y} & {k_y^2  - k_x^2 }  \\
\end{array}} \right] + U(\bm{\rho})\hat I \:.
\end{equation}
Here $\bm{k}=(k_x,k_y)$ is the momentum operator divided by the reduced Planck
constant $\hbar$, $\gamma$ and $\gamma^\prime$ are the band parameters which
are expressed in terms of the parameters $L$ and $M$ (see
Ref.~[\onlinecite{BirPikus}]) 
% (see Refs.~[\onlinecite{BirPikus,Cardona}])
via $\gamma=(L+M)/2$ and
$\gamma^\prime=(L-M)/2$, $U(\bm{\rho})$ is the attractive potential of the
acceptor, and $\hat I$ is the unit matrix $2 \times 2$. We assume the acceptor
potential $U(\bm{\rho})$ to be isotropic. Then, the hole states are described by 
the quantum number $n$ ($n=0,1,2, \ldots$) and the projection of the orbital angular momentum onto the $z$ axis $j$ ($j=0,\pm1,\pm2, \ldots$). The corresponding
wave functions $\psi_{n j}(\bm{\rho})$ in the polar coordinates
$\bm{\rho}=(\rho,\varphi)$ are given by
\begin{equation}\label{eqPsi}
\psi_{n j}(\bm{\rho})   = \frac{{e^{i\left( {j
- 1} \right)\varphi } }}{{\sqrt 2 }} \left[ {\begin{array}{*{20}c}
   \eta_{n j}(\rho) + \xi_{n j}(\rho) e^{2i\varphi}  \\
   i \eta_{n j}(\rho) - i \xi_{n j}(\rho) e^{2i\varphi}
     \\
\end{array}} \right] \:,
\end{equation}
where $\eta_{n j}(\rho)$ and $\xi_{n j}(\rho)$ are solutions of the equation set 
\begin{widetext}
\begin{eqnarray}\label{eqSystem}
\gamma \left[ \frac{d^2}{d \rho^2}  + \frac{1}{\rho} \frac{d}{d \rho} -
\frac{(j+1)^2}{\rho^2} \right] \xi_{n j}(\rho) + \gamma^\prime \left[
\frac{d^2}{d\rho^2} + \frac{1-2j}{\rho} \frac{d}{d \rho} + \frac{j^2
-1}{\rho^2} \right] \eta_{n j}(\rho) = [U(\rho) - E_{n j}] \xi_{n j}(\rho) \:, \\
\gamma^\prime \left[ \frac{d^2}{d\rho^2} + \frac{1+2j}{\rho} \frac{d}{d \rho} +
\frac{j^2 -1}{\rho^2} \right] \xi_{n j}(\rho) + \gamma \left[ \frac{d^2}{d \rho^2} +
\frac{1}{\rho} \frac{d}{d \rho} - \frac{(j-1)^2}{\rho^2} \right] \eta_{n j}(\rho) =
[U(\rho) - E_{n j}] \eta_{n j}(\rho) \:, \nonumber
\end{eqnarray}
\end{widetext}
$E_{n j}$ is the energy of the state $\psi_{n j}(\bm{\rho})$ measured from the subband edge. We note that the states $(n,j)$ and $(n,-j)$ have the same energy and their envelope functions are related by $\psi_{n,-j}(\bm{\rho})=\psi_{n,j}^*(\bm{\rho})$. From Eq.~(\ref{eqPsi}) it follows that only the states with $j=\pm1$ contain envelope functions of $s$-type which are nonzero at the acceptor. Therefore, the ground state of localized holes is formed with the functions $\psi_{0,\pm1}(\bm{\rho})$. Below we denote them as $\psi_{\pm}(\bm{\rho})$, respectively.

In the absence of strain, the ground state is degenerate in the projection of angular momentum and the hole wave function represents a superposition of $\psi_{+}(\bm{\rho})$ and $\psi_{-}(\bm{\rho})$. The acoustic soliton lifts the degeneracy causing  the transition of a hole between the above states.  

\begin{figure}[t] 
 \leavevmode
% \centering\epsfxsize=220pt \epsfbox[100 350 600 670]{fig2.eps}
 \centering\includegraphics[width=0.47\textwidth]{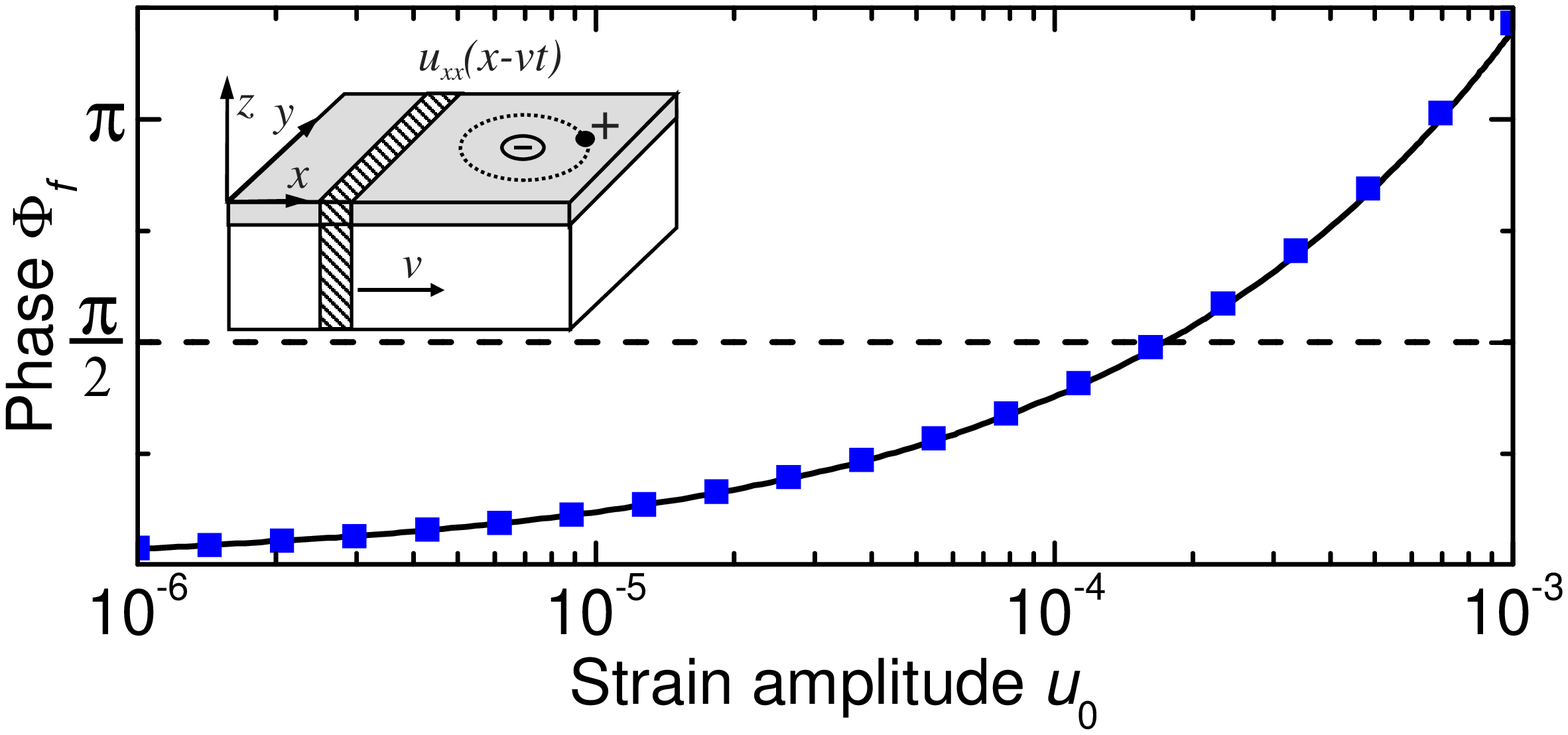}
 \caption{(Color online) Dependence of the phase shift $\Phi_f$ on the strain amplitude $u_0$ calculated numerically for the Coulomb potential (squares) and plotted after Eq.~(\ref{eqPhi2Result}) for the zero-radius potential (solid curve).
Dashed line corresponds to $\Phi_f=\pi/2$ which is optimal for the state conversion. Inset sketches the propagation of strain soliton along the quantum well.}
 \label{fig_phase}
\end{figure}

The strain effect on the heavy-hole subband is described by the effective Hamiltonian~\cite{BirPikus}
\begin{equation}\label{eqV2} 
V = \left[ {\begin{array}{*{20}c}
   l u_{xx} + m u_{yy} & n u_{xy}  \\
   n u_{xy} & m u_{xx} + l u_{yy}   \\
\end{array}} \right]  \:,
\end{equation}
where $l$, $m$, and $n$ are the deformation potential constants, $u_{\alpha\beta}$ are the strain tensor components. We consider that the bulk acoustic soliton propagates in the $x$ direction along the quantum well plane inducing time-dependent component  $u_{xx}(x,t)$ of the strain tensor, see inset to Fig.~\ref{fig_phase}. The strain pulse is assumed weak enough not to cause ionization of the localized hole or its transition to excited states. Then, according to the perturbation theory, the hole wave function $\psi(\bm{\rho},t)$ can be expanded over the non-disturbed states $\psi_{\pm}(\bm{\rho})$ as follows
\begin{equation}\label{eqPsic1c2} 
\psi(\bm{\rho},t) = c_{+}(t) \psi_{+}(\bm{\rho}) + c_{-}(t) \psi_{-}(\bm{\rho}),
\end{equation}
where the coefficients $c_{j}(t)$ ($j=\pm$) satisfy the coupled equations
\begin{eqnarray}\label{eqSystemc1tc2t} 
i\hbar \frac{d c_{+}(t)}{d t} =  V_{++}(t) \,c_{+}(t) +  V_{+-}(t)\,c_{-}(t) \:, \\
i\hbar \frac{d c_{-}(t)}{d t} = V_{-+}(t) \,c_{+}(t) +  V_{--}(t)\,c_{-}(t) \:, \nonumber
\end{eqnarray}
and $V_{jj'}(t)=\int \psi_{j}^{\dag}(\bm{\rho}) V \psi_{j'}(\bm{\rho}) d \bm{\rho}$ are the matrix elements of the perturbation~(\ref{eqV2}). The specific form of the functions~(\ref{eqPsi}) and the perturbation $V \propto u_{xx}(x,t)$ leads to the relations $V_{++}(t)=V_{--}(t)$ and $V_{+-}(t)=V_{-+}(t)$, which makes Eqs.~(\ref{eqSystemc1tc2t}) easily solvable.

The solution of Eqs.~(\ref{eqSystemc1tc2t}) assumes the form
\begin{eqnarray}\label{eqc1c2}
c_{+}(t)= \left[a_{+}\cos{\Phi(t)} - i a_{-}\sin{\Phi(t)} \right] {\rm e}^{-i\Theta(t)} \:, \\
c_{-}(t)= \left[a_{-}\cos{\Phi(t)} - i a_{+}\sin{\Phi(t)} \right] {\rm e}^{-i\Theta(t)} \:, \nonumber
\end{eqnarray}
where $a_{+}=c_{+}(-\infty)$ and $a_{-}=c_{-}(-\infty)$ are coefficients of the initial state at $t=-\infty$, $\Phi(t)$ and $\Theta(t)$ are phases given by
\begin{equation}\label{phi_t}
\Phi(t)=\frac{1}{\hbar}\int\limits_{ - \infty}^t {V_{+-}(t')dt'} \:, \;\;
\Theta(t)=\frac{1}{\hbar}\int\limits_{ - \infty}^t {V_{++}(t')dt'} \:.
\end{equation}   
It follows from Eqs.~(\ref{eqc1c2}) that it is the phase $\Phi(t)$ that describes the hole state evolution while $\Theta(t)$ constitutes the common phase factor. In the final state, i.e., at $t=+\infty$ 
when the soliton has completely passed through the area of hole localization, the phase shift has the form
\begin{equation}\label{DeltaPhi}
\Phi_f  = \frac{l-m}{2\hbar} \int \eta_{0,1}^2(\rho) d \bm{\rho}
\int_{-\infty}^{+\infty} u_{xx}(x,t) dt \:.
\end{equation}
Since the average value of the strain $u_{xx}$ is non-zero within the soliton, 
$\Phi_f \neq 0$, and the final hole state differs from the initial one. The most efficient conversion of the hole state occurs if the soliton amplitude is so that  
\begin{equation}
|\Phi_f| = \pi(n + 1/2), \;\;\;n=0,1,2,...
\end{equation}
In this particular case, the soliton-hole interaction results in the complete switching of the hole angular momentum: The initial state $\psi_{+}(\bm{\rho})$ is converted into the state $\psi_{-}(\bm{\rho})$ and vice versa, see Eq.~(\ref{eqc1c2}).

To elaborate the effect of state switching we consider two models of the acceptor: (i) Coulomb potential $U(\rho)=-e^2/(\epsilon\rho)$, where $e$ is the elementary charge and $\epsilon$ is the dielectric constant, and (ii) zero-radius potential.~\cite{Averkiev04,Monakhov06}
In the first case, the ground hole states and the localization energy $E_0>0$ are calculated numerically by solving Eqs.~(\ref{eqSystem}). The latter approach allows us to find the states analytically for a given $E_0$ assuming that the range of attractive potential is much shorter than the hole localization length. Within this approach, solution of Eq.~(\ref{eqSystem}) for the ground state has the form 
\begin{eqnarray}\label{Psi_zr}
\eta_{0,1}(\rho) = \frac{\rho_1\rho_2}{\sqrt{2\pi(\rho_1^2+\rho_2^2)}} \left[ \frac{K_0(\rho/\rho_1)}{\rho_1^2} + \frac{K_0(\rho/\rho_2)}{\rho_2^2} \right] \:, \\
\xi_{0,1}(\rho) = \frac{\rho_1\rho_2}{\sqrt{2\pi(\rho_1^2+\rho_2^2)}} \left[ \frac{K_2(\rho/\rho_1)}{\rho_1^2} - \frac{K_2(\rho/\rho_2)}{\rho_2^2} \right] \:, \nonumber
\end{eqnarray}
where $K_0(x)$ and $K_2(x)$ are the Macdonald functions, the radii $\rho_1$ and $\rho_2$ are given by $\rho_1=\sqrt{-(\gamma+\gamma')/E_0}$, $\rho_2=\sqrt{-(\gamma-\gamma')/E_0}$, and it is assumed that $\gamma+\gamma', \gamma-\gamma' <0$.

Acoustic solitons in crystals are typically approximated by solutions of the Korteweg--de Vries (KdV) wave equation~\cite{Hao01} or the doubly dispersive equation (DDE).~\cite{Khusnutdinova08} Since the particular soliton 
shape is not crucial for our results, we consider the KdV soliton so that $u_{xx}$ has the form
\begin{equation}\label{eqCosh} 
u_{xx}(x,t) = u_0\cosh ^{-2}
\left(\frac{{x - vt}}{L}\right),
\end{equation}
where $u_0$ is the strain amplitude, $v$ and $L=d/\sqrt{u_0}$ are the soliton velocity and size, respectively, and  $d$ is a material constant. In this case 
$\int_{-\infty}^{+\infty} u_{xx}(x,t) dt = 2d\sqrt{u_0}/v$.

Figure~\ref{fig_phase} shows dependence of the phase shift $\Phi_f$ on the strain
amplitude $u_0$. 
Squares correspond to the numerical calculation of $\Phi_f$ given by Eq.~(\ref{DeltaPhi}) for the Coulomb localizing potential, solid curve represents the analytical dependence
\begin{equation}\label{eqPhi2Result} 
\Phi_f = \frac{{l - m}}{{2\hbar }}\frac{{d
\sqrt{u_0}}} {v} \left[ {1 + \frac{{\gamma^2  - \gamma^{\prime 2}
}}{{2\gamma \gamma^\prime }}\ln \left( {\frac{{\gamma + \gamma^\prime
}}{{\gamma  - \gamma^\prime }}} \right)} \right]
\end{equation}
derived in the zero-radius-potential approach by integrating the wave function~(\ref{Psi_zr}). The band parameters of silicon used in the calculation are as follows: $\gamma = -4.65$ $\hbar/(2m_0)$, $\gamma^\prime = -1.13$ $\hbar/(2m_0)$,~\cite{IP_book} where $m_0$ is free electron mass, the deformation-potential constants $l=-4.9$~eV and $m=-1.5$~eV.~\cite{BirPikus}
The soliton parameters $d=1.5$~\AA, $v=0.9\cdot10^{6}$~cm/s, and the strain amplitude $u_0$ in the range $10^{-6} \div 10^{-3}$ are chosen, which corresponds to experimental data on acoustic phonon pulses in silicon.~\cite{Daly04} 

From Fig.~\ref{fig_phase} it follows that the phase shift $\Phi_f=\pi/2$ 
%(shown by the dashed horizontal line)
optimal for the state conversion is achieved in Si-based structures at a moderate strain amplitude $u_0 \approx 1.8\cdot 10^{-4}$. It proves that the hole states can be efficiently manipulated by acoustic solitons. Moreover, both Coulomb and zero-radius potentials lead to the same quantitative result indicating that $\Phi_f$ weakly depends on the form of localizing potential for silicon band parameters. This can be attributed to the fact that the envelope function of the ground heavy-hole state is mainly of $s$-type even at $\gamma^\prime/\gamma \approx 0.24$. Therefore, the integral $\int \eta_{0,1}^2(\rho) d \bm{\rho}$ determining the conversion efficiency [see Eq.~(\ref{DeltaPhi})] is approximately equal to 1 in accordance with the wave function normalization and independent of the explicit form of $\eta_{0,1}(\rho)$.

For the strain amplitude $u_0=1.8 \cdot 10^{-4}$ optimal for the state conversion, the soliton length $L$ is estimated as $110$~\AA\mbox{} that is approximately 7 times larger than the radius of hole localization $a_0 = e^2/(\epsilon E_0) \approx 16$~\AA\mbox{} in silicon quantum wells. In the approximation of $L \gg a_0$, the time evolution of the phase $\Phi(t)$ for the soliton shape~(\ref{eqCosh}) has the form
\begin{equation}
\Phi(t) = \Phi_f \frac{\tanh{(vt/L)}+1}{2}  \:.
\end{equation}

\begin{figure}[h]
 \centering\includegraphics[width=0.45\textwidth]{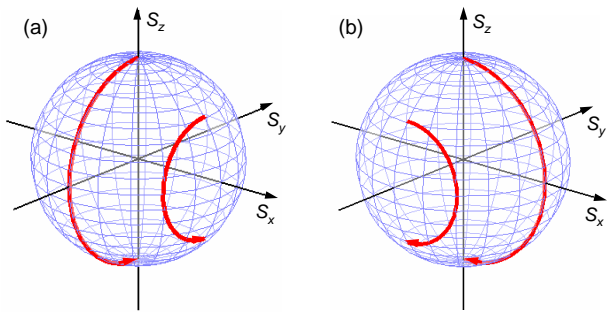}
 \caption{(Color online) Trajectories on the Bloch sphere corresponding to the time evolution of the hole state caused by acoustic soliton propagating along (a) $x$ and (b) $y$ axes.}
 \label{fig_sphere}
\end{figure}

The soliton-induced evolution of a hole state can fruitfully be considered as a precession of
pseudospin or a trajectory on the Bloch sphere. In this approach,~\cite{Feynman57} any superposition
~(\ref{eqPsic1c2}) of two states, $\psi_{+}(\bm{\rho})$ and $\psi_{-}(\bm{\rho})$, is attributed to a unit vector $\bm{S}$ or a point $(S_x,S_y,S_z)$ on the sphere, see Fig.~\ref{fig_sphere}. The components of $\bm{S}$ are given by
\begin{equation}\label{pseudospin}
\bm{S} = \chi^\dag \, \bm{\sigma} \chi \:,
\end{equation}
where $\bm{\sigma}$ is the vector of Pauli matrices and $\chi$ is the spinor composed of the coefficients $c_{+}(t)$ and $c_{-}(t)$, $\chi = [c_{+}(t),c_{-}(t)]^T$. In particular, the pure states $\psi_{+}(\bm{\rho})$ and $\psi_{-}(\bm{\rho})$ correspond to the polar points $(0,0,1)$ and $(0,0,-1)$, respectively, while the states 
$[\psi_{+}(\bm{\rho}) + e^{i \alpha} \psi_{-}(\bm{\rho})]/\sqrt{2}$ with arbitrary phase $\alpha$ are projected onto the Bloch sphere equator. 

During the soliton-hole interaction the coefficients $c_{\pm}$ vary in time, see Eq.~(\ref{eqc1c2}), which corresponds to motion of the point $(S_x,S_y,S_z)$ on the Bloch sphere. Equation describing the time evolution of $\bm{S}$ can be derived from Eq.~(\ref{eqSystemc1tc2t}). It gives
\begin{equation}\label{dSst}
\frac{d \bm{S}}{dt} = [\bm{\Omega}(t) \times \bm{S}] \:,
\end{equation}
where $\bm{\Omega}(t) = (1/\hbar) \sum_{jj'} V_{jj'}(t) \,\bm{\sigma}_{j'j}$. For a given initial state $\bm{S}_0$ and $\bm{\Omega}(t)$, Eq.~(\ref{dSst}) allows one to calculate the trajectory $\bm{S}(t)$ and restore the wave function. 
Shown in Fig.~\ref{fig_sphere} are examples of such trajectories under a soliton-induced uniaxial strain. 
The strain $u_{xx}(x,t)$ produced by soliton propagating along the $x$ axis corresponds to $\bm{\Omega}(t) \parallel x$ and, therefore, causes $\bm{S}$ to rotate around the $x$ axis (Fig.~\ref{fig_sphere}a) while the strain $u_{yy}(y,t)$ corresponds to $\bm{\Omega}(t) \parallel y$ leading to the rotation of $\bm{S}$ around the $y$ axis (Fig.~\ref{fig_sphere}b). In the case of complete conversion of the hole state from $\psi_{+}(\bm{\rho})$ into $\psi_{-}(\bm{\rho})$, the trajectories pass from one pole of the Bloch sphere to the other.

The time of switching the hole states is given by $\tau=2(L+a_0)/v$. It is of nanosecond scale and much less than the lifetime of localized holes at low temperatures as well as the soliton lifetime. It suggests that the effect can be studied experimentally by means of time- and space-resolved optical spectroscopy. Indeed, the states $\psi_{+}(\bm{\rho})$ and $\psi_{-}(\bm{\rho})$ are characterized by projections $+1$ and $-1$ of the orbital angular momentum, therefore, optical pumping by circularly polarized light leads to a predominant population of one of the states. The hole angular momentum can be registered in turn by analyzing
the polarization of recombination radiation or the polarization change of a probe pulse (see, e.g., Ref.~[\onlinecite{Hubner08}]). The soliton propagation through the area of hole localization reverses the projection of hole angular momentum leading to a change in optical response.

To summarize, we have demonstrated the possibility of switching the quantum-mechanical state of a localized hole by an acoustic soliton. The amplitude of the strain soliton required for switching the hole states in silicon-based quantum wells is found to be $\approx 2\cdot10^{-4}$ which corresponds to strain pulses studied experimentally. 

This work was supported by the Russian Foundation for Basic Research, Programs of the RAS and ``Leading Scientific Schools'' (grants 1972.2008.2 and 3415.2008.2).


\begin{thebibliography}{99}
%\itemsep-2pt
\bibitem{BirPikus} G.~L.~Bir and G.~E.~Pikus, \textit{Symmetry and
Strain-Induced Effects in Semiconductors} (Wiley, New York, 1974).

\bibitem{Sun07} Y.~Sun, S.~E.~Thompson, and T.~Nishida,
Physics of strain effects in semiconductors and metal-oxide-semiconductor
field-effect transistors,
J. Appl. Phys. {\bf 101}, 104503 (2007).

\bibitem{Hao01}
H.-Y.Hao, H.J. Maris,
Experiments with acoustic solitons in crystalline solids,
Phys. Rev. B \textbf{64} 064302 (2001).

\bibitem{Muskens04} O.~L.~Muskens, A.~V.~Akimov, and J.~I.~Dijkhuis,
Coherent interactions of terahertz strain solitons and electronic two-level systems in photoexcited ruby,
Phys. Rev. Lett. {\bf 92}, 035503 (2004).

%\bibitem{Muskens05}
%O.~L.~Muskens, J.~I.~Dijkhuis,
% Interactions of ultrashort strain solitons and terahertz electronic two-level systems in photoexcited ruby
%Phys. Rev. B \textbf{71}, 104304 (2005).

\bibitem{Scherbakov07}
A.~V.~Scherbakov, P.~J.~S.~van~Capel, A.~V.~Akimov, J.~I.~Dijkhuis,
D.~R.~Yakovlev, T.~Berstermann, M.~Bayer,
Chirping of an optical transition by an ultrafast acoustic soliton train in a semiconductor quantum well,
Phys. Rev. Lett, \textbf{99}, 057402 (2007).

\bibitem{Voronkov99} E.~N.~Voronkov,
Detection of the decay of an acoustic soliton arising during pulsed breakdown of a glassy semiconductor film,
JETP Lett. {\bf 70}, 72 (1999).

\bibitem{Kosevich} A.~M.~Kosevich, \textit{Theory of Crystal Lattice} (WILEY-VCH, Berlin,
New York, 1999).

\bibitem{Rodriguez99} S.~Rodr\'{i}guez, J.~A.~L\'{o}pez-Villanueva, I.~Melchor,
and J.~E.~Carceller,
Hole confinement and energy subbands in a silicon inversion layer using the
effective mass theory,
J. Appl. Phys. {\bf 86}, 438 (1999).

%\bibitem{Cardona} P.~Y.~Yu and M.~Cardona, \textit{Fundamentals of Semiconductors}
%(Springer, Berlin, 1996).

%A.~M.~Samsonov, Strain Solitons in Solids and How to Construct Them,
%Chapman and Hall/CRC, Boca Raton, FL, 2001.
\bibitem{Khusnutdinova08} K.~R.~Khusnutdinova and A.~M.~Samsonov, 
Fission of a longitudinal strain solitary wave in a delaminated bar,
Phys. Rev. E \textbf{77}, 066603 (2008).

\bibitem{Averkiev04} N.~S.~Averkiev, A.~E.~Zhukov, Yu.~L.~Ivanov,
P.~V.~Petrov, K.~S.~Romanov, A.~A.~Tonkikh, V.~M.~Ustinov, and
G.~E.~Zyrlin, 
Energy structure of A(+) centers in quantum wells,
Semicond. \textbf{38}, 217 (2004).

\bibitem{Monakhov06} A.~M.~Monakhov, K.~S.~Romanov, I.~E.~Panaiotti, and N.~S.~Averkiev,
Spatial distribution of a hole localized on a magnetic acceptor in cubic crystals,
Solid State Commun., \textbf{140}, 422 (2006).

\bibitem{IP_book} E. L. Ivchenko and G. Pikus, \textit{Superlattices and Other 
Heterostructures. Symmetry and Optical Phenomena} (Springer, Berlin, 1997).

\bibitem{Daly04} B.~C.~Daly, T.~B.~Norris, J.~Chen, J.~B.~Khurgin, 
Picosecond acoustic phonon pulse propagation in silicon,
Phys. Rev. B \textbf{70}, 214307 (2004).

\bibitem{Feynman57} R.~P.~Feynman, F.~L.~Vernon, and R.~W.~Hellwarth,
Geometrical representation of the Schr\"{o}dinger equation for solving maser problems,
J. Appl. Phys. {\bf 28}, 49 (1957).

\bibitem{Hubner08} J.~H\"{u}bner and M.~Oestreich, 
Time-resolved spin dynamics and spin noise spectroscopy,
in \textit{Spin Physics in Semiconductors}, edited by M~I.~Dyakonov (Springer, Berlin, 2008).

\end{thebibliography}
\end{document}